\documentclass[prl,twocolumn,superscriptaddress,showpacs]{revtex4}

\usepackage[german,english]{babel}
\usepackage{amssymb}
\usepackage{amsfonts}

\begin{document}

\title{Fractional Klein-Kramers equation for superdiffusive transport:
normal versus anomalous time evolution in a differential L{\'e}vy walk
model.}

\date{\today}

\author{Ralf Metzler}
\affiliation{Department of Physics, Massachusetts Institute of Technology,
77 Massachusetts Avenue, Room 12-109, Cambridge, Massachusetts 02139, USA}
\author{Igor M. Sokolov}
\affiliation{Theoretische Polymerphysik, Universit\"at
Freiburg, Hermann-Herder-Str.\ 3, 79104 Freiburg i.Br., Germany}

\begin{abstract}
We introduce a fractional Klein-Kramers equation which describes sub-ballistic
superdiffusion in phase space in the presence of a space-dependent external
force field. This equation defines the differential L{\'e}vy walk model whose
solution is shown to be non-negative. In the velocity coordinate, the
probability density relaxes in Mittag-Leffler fashion towards the Maxwell
distribution whereas in the space coordinate, no stationary solution exists
and the temporal evolution of moments exhibits a competition between
Brownian and anomalous contributions.
\end{abstract}


\maketitle

Classically, Brownian stochastic transport processes in the phase space
spanned by velocity $v$ and coordinate $x$ are described by the deterministic
Klein-Kramers equation (KKE). In the low and high friction limits, the KKE
reduces to the Rayleigh equation which describes the relaxation of the
velocity probability density function (pdf) towards the Maxwell distribution,
and the Fokker-Planck-Smoluchowski equation controlling the temporal approach
of the Gibbs-Boltzmann equilibrium, respectively \cite{klein,vankampen}. The
KKE is therefore a fundamental equation in non-equilibrium systems dynamics.

Brownian transport is characterised through the Gaussian
pdf and the linear time dependence $\langle x^2(t)
\rangle=2Kt$ of the mean squared displacement in the force-free diffusion
limit, its universality being guaranteed by the central limit theorem
\cite{vankampen}. In a broad variety of systems, however, it has been
found that correlations in space or time give birth to {\em anomalous}
transport whose pdf is non-Gaussian and/or whose mean squared
displacement is non-linear in time \cite{bouchaud}. These systems
include charge carrier transport in amorphous
semiconductors \cite{pfister}, tracer dispersion in convection rolls and
rotating flows \cite{ypp,swinney}, capillary surface waves \cite{hansen},
the motion of bacteri{\ae} and the flight of an albatross \cite{levandowsky},
intracellular transport \cite{granek}, transport in micelles \cite{ott}, 2-D
dusty plasmas \cite{juan}, the dynamics in (bio)polymeric systems
\cite{amblard}, and the NMR diffusometry in porous glasses and percolation
clusters \cite{sigi}, among others.

The systems we are interested in fall into the broad class whose {\em
force-free} diffusion behaviour is characterised through the power-law
form $\langle x^2 (t)\rangle\propto t^{\varkappa}$, which separates into
subdiffusion ($0<\varkappa<1$) and superdiffusion ($\varkappa>1$). The
continuous time random walk (CTRW) model has proved to be a well-suited
framework which accounts for such anomalous diffusion for the entire spectrum
of $\varkappa$ \cite{klablushle}.
Especially in the sub-ballistic superdiffusive domain $1<\varkappa<2$,
L{\'e}vy walks which couple long flight times with a time cost have been
a successful tool, e.g., in fluid dynamics \cite{swinney,fluid}. The
space-time coupling of L{\'e}vy walks leads to finite moments and they are
therefore fundamentally different from L{\'e}vy flights which exhibit a
diverging variance \cite{report}.

In the presence of external force fields, the CTRW approach is less flexible.
It has been realised that fractional equations constitute a tailor-made
framework to formulate the underlying dynamics equations in coordinate and
phase space; see, for instance, \cite{kusnezov,lutz,mebakla,report,eli} and
references therein. In the subdiffusive domain $0<\varkappa<1$, a fractional
KKE was derived from CTRW models \cite{mebakla1} and from the
Chapman-Kolmogorov equation \cite{gcke}. A similar consistent generalisation
to systems in the regime $1<\varkappa<2$ is still outstanding. In this note,
we propose the fractional KKE
\begin{widetext}
\begin{equation}
\label{fkk}
\frac{\partial P}{\partial t}=\left(-\frac{\partial}{\partial x}v-
\, _0D_t^{1-\alpha}\varrho_{\alpha}\left[-\gamma\frac{\partial}{
\partial v}v+\frac{F(x)}{m}\frac{\partial}{\partial v}-\kappa\frac{
\partial^2}{\partial v^2}\right]\right)P(x,v,t), \,\, 0<\alpha<1
\end{equation}
\end{widetext}
for the description of sub-ballistic superdiffusive anomalous transport. In Eq.
(\ref{fkk}), $P(x,v,t)dxdv$ is the joint probability to find the test particle
of mass
$m$ with coordinate $x,\ldots,x+dx$ and velocity $v,\ldots,v+dv$ at time $t$.
$\gamma$ denotes the friction constant which quantifies the effective
dissipative interaction with the environment, $\kappa$ is the velocity
diffusion constant and $F(x)=-d\Phi(x)/dx$ is an external force field.
The factor $\varrho_{\alpha}$ has dimension $[\varrho_{\alpha}]={\rm sec}^{
-\alpha}$ and is a function of some time scale $\tau$ characteristic for
the waiting time process and of an interaction time scale $\tau^*$
\cite{gcke}, see also below. In the limit $\alpha=1$, Eq. (\ref{fkk})
corresponds to the
standard KKE whereas for $0<\alpha<1$, the process contains a scale-free
memory of the power-law type entering
through the fractional Riemann-Liouville operator $_0D_t^{1-\alpha}
\equiv\partial/\partial t\big(\,_0D_t^{-\alpha}\big)$ with
\cite{oldham}
\begin{equation}
\label{rl}
_0D_t^{-\alpha}W(v,t)\equiv\int_0^t dt' W(v,t')(t-t')^{\alpha-1}/
\Gamma(\alpha).
\end{equation}
The operator $_0D_t^{-\alpha}$ possesses the important property $\int_0
^{\infty}e^{-ut}\,_0D_t^{-\alpha}f(t)=u^{-\alpha}f(u)$ under Laplace
transformation.

With respect to its coordinate $x$, Eq. (\ref{fkk}) is a Liouville-type
equation which gives rise to the relation $\frac{d}{dt}\langle\langle x(t)
\rangle\rangle=\langle\langle v(t)\rangle\rangle$ between the velocity- and
coordinate-averaged random variables $x$ and $v$; by this, the KKE (\ref{fkk})
differs from the subdiffusive model in Ref. \cite{gcke}. The action of the
friction, force and velocity diffusion in Eq. (\ref{fkk}) enters through the
non-local memory relation brought about by the fractional operator; i.e., the
local fluctuations of velocity which follow from a Fokker-Planck-like equation
in the Brownian case are now described by a fractional Fokker-Planck equation.
The connection to the space coordinate is assumed to be
given through the classical drift term $-\partial vP/\partial x$. From these
basic physical assumptions, the fractional KKE (\ref{fkk}) emerges.

Let us explore the behaviour predicted by Eq. (\ref{fkk}) in more detail. By
integration over the spatial coordinate $x$, the fractional equation for the
velocity pdf $P(v,t)$
\begin{equation}
\label{fr}
\frac{\partial P}{\partial t}=\, _0D_t^{1-\alpha}\varrho_{\alpha}
\left(\gamma\frac{\partial}{\partial v}v-\frac{F(x)}{\gamma m}\frac{
\partial}{\partial v}+\kappa\frac{\partial^2}{\partial
v^2}\right)P(v,t)
\end{equation}
obtains which reduces to its Brownian counterpart in the limit $\alpha=1$.

In the force-free limit, (\ref{fr}) corresponds to the fractional Rayleigh
equation discussed in Refs. \cite{gcke,basil}. The relaxation of velocity
moments can be inferred directly from Eq. (\ref{fr}) in the force-free case. 
Accordingly, the fractional relaxation equation \cite{report,wg}
\begin{equation}
{\textstyle\frac{d}{dt}}\langle v(t)\rangle=-\gamma_{\alpha}
\,_0D_t^{1-\alpha}\langle v(t)\rangle
\end{equation}
emerges where $\gamma_{\alpha}\equiv\varrho_{\alpha}\gamma$. For $v_0=v(0)$,
its solution
\begin{equation}
\langle v(t)\rangle=v_0E_{\alpha}\left(-\gamma_{\alpha} t^{\alpha}\right)
\end{equation}
features the Mittag-Leffler function $E_{\alpha}\left(-\gamma_{\alpha}
t^{\alpha}\right)\equiv\sum_{n=0}^{\infty}\left(-\gamma_{\alpha} t^{\alpha}
\right)^n/\Gamma(1+\alpha n)$ which is monotonically decaying and which
interpolates between the initial stretched exponential behaviour $E_{\alpha}
\left(-\gamma_{\alpha} t^{\alpha}\right)\sim\exp \left(-\gamma_{\alpha}
t^{\alpha}/\Gamma(1+\alpha)\right)$ and the final inverse power-law pattern
$E_{\alpha}\left(-\gamma_{\alpha} t^{\alpha}\right)\sim\left(\gamma_{\alpha}
t^{\alpha}\Gamma(1-\alpha)\right)^{-1}$ \cite{erdelyi}. Thus,
the mean relaxation time diverges. The second moment
\begin{equation}
\langle v^2(t)\rangle=\kappa/\gamma+(v_0^2-\kappa/\gamma)E_{\alpha}\left(
-2\gamma_{\alpha} t^{\alpha}\right)
\end{equation}
equilibrates Mittag-Leffler fashion towards the equilibrium value $\kappa/
\gamma$, and the pdf $P(v,t)$ relaxes towards a Gaussian. In
thermodynamical systems, by comparison, the velocity pdf follows
the Maxwell distribution $P(v)=\left(2\pi k_BT/m\right)^{-1/2}\exp\left(
-mv^2/[2k_BT]\right)$, so that we find the Einstein relation $\kappa=k_BT
\gamma/m$ \cite{report,eli}. The velocity-velocity correlation function
associated with Eq. (\ref{fr}) is
\begin{equation}
\langle v(0)v(t)\rangle=v_0^2E_{\alpha}\left(-\gamma_{\alpha}t^{\alpha}\right)
\sim v_0^2\left(\gamma_{\alpha}t^{\alpha}\right)^{-1}
\end{equation}
whose long time behaviour is equivalent to the CTRW-L{\'e}vy walk result
where the $t^{-\alpha}$-scaling follows from the behaviour of the cumulative
waiting time distribution.

This corresponds to the physical model underlying Eq. (\ref{fkk}):
in the velocity space, we assume that the particle undergoes
collisions, i.e., random velocity changes, such that successive collisions
are separated by time spans governed through the long-tailed waiting time
pdf $\psi({\mathfrak t})\sim A_{\alpha}({\mathfrak t}/\tau)^{-1-\alpha}$
whose characteristic time ${\mathfrak T}=\int_0^{\infty}{\mathfrak t}\psi(
{\mathfrak t})d{\mathfrak t}$ diverges. In this case, the velocity-velocity
correlation function is proportional to the probability that no scattering
took place before time $t$, and thus $\left\langle v(0)v(t)\right\rangle =
v_0^2\int_t^{\infty}\psi(t)dt\propto v_0^2(t/\tau)^{-\alpha}$. In addition,
we assume that the external force field $F(x)$ constantly acts on the test
particle during its sojourns.

The characteristic behaviour of this model is that it includes arbitrarily
long sojourns but leads to finite moments of any order and
an exponentially decaying pdf $P$. We call the process associated with the
fractional KKE (\ref{fkk}) a differential L{\'e}vy walk model noting that
it is an approximation to CTRW-L{\'e}vy walks which offers the distinct
possibility to study the effects of an external force field on the behaviour
of lower order moments of a L{\'e}vy walk.

The velocity average of the fractional KKE
(\ref{fkk}) can be performed by integration over $\int dv$ and $\int
vdv$, and combination of the two resulting equations. With $\langle v^2\rangle
=k_BT/m$, this procedure yields the fractional telegrapher's type equation
\begin{equation}
\frac{1}{\gamma_{\alpha}}\frac{\partial^2P}{\partial t^2}+\,_0D_t
^{2-\alpha}P=\left(-\,_0D_t^{1-\alpha}\frac{\partial}{\partial x}
\frac{F(x)}{\gamma m}+K\frac{\partial^2}{\partial x^2}\right)P
(x,t)
\end{equation}
which exhibits a transition from a short-time ballistic
behaviour with $\langle x^2(t)\rangle\sim(K\gamma)t^2$ in the force-free case,
to the long-time or high-friction limit governed through the
fractional Fokker-Planck-Smoluchowski equation
\begin{equation}
\label{ffp}
\frac{\partial P}{\partial t}=\left(-\frac{\partial}{
\partial x}\frac{F(x)}{\gamma m}+\,_0D_t^{\alpha-1}K\frac{\partial^2}
{\partial x^2}\right)P(x,t).
\end{equation}
Its force-free limit exactly corresponds to the enhanced diffusion approach
reported in Ref. \cite{abm} which exhibits a distinct bimodal behaviour
characterised through two separating humps. From Eq. (\ref{ffp}) one observes
the difference to the fractional Fokker-Planck equation for subdiffusive
processes from Ref. \cite{mebakla}: here, the force is temporally local with
$\partial P/\partial t$. In Eq. (\ref{ffp}), we identify $K\equiv k_BT/(m
\gamma_{\alpha})$; note that this does not represent a real Einstein
relation as Eq. (\ref{ffp}) does not have a stationary solution, see
below. The mean squared displacement corresponding to Eq. (\ref{ffp})
with $F(x)=0$ is given through
\begin{equation}
\label{msd}
\langle x^2(t)\rangle=2Kt^{2-\alpha}/\Gamma(3-\alpha)
\end{equation}
which describes sub-ballistic superdiffusion. This result is equivalent to
the fractional Kramers model discussed in Ref. \cite{basil}, and to the
CTRW-L{\'e}vy walk \cite{klablushle}. It is in the presence of an external
force that the fractional KKE (\ref{fkk}) exhibits a fundamentally different
behaviour.

To see this, consider the fractional Fokker-Planck equation (\ref{ffp}) for
non-trivial types of the external force $F(x)$. Accordingly, the first
moment is given by $\frac{d}{dt}\langle x(t)\rangle=\langle F(x)\rangle/(
m\gamma)$ which can be solved for constant or linear forces. In particular,
for the constant drift $F(x)=Vm\gamma$, the first moment becomes 
\begin{equation}
\langle x(t)\rangle=Vt
\end{equation}
which corresponds to the traditional drift behaviour \cite{vankampen}. The
second moment
\begin{equation}
\langle x^2(t)\rangle=V^2t^2+2Kt^{2-\alpha}/\Gamma(3-\alpha)
\end{equation}
combines this drift with the sub-ballistic behaviour $\propto t^{2-\alpha}$
such that the variance $\langle\Delta x(t)^2\rangle\equiv(\langle x^2(t)
\rangle-\langle x(t)\rangle^2)$ is given by Eq. (\ref{msd}). This behaviour
is analogous to the Galilei invariant diffusion-advection model derived in
Ref. \cite{igor} for the subdiffusive case. In our case, the analytical
solution is given by the free (superdiffusive) solution explored in detail
in Ref. \cite{abm}, taken
at the translated coordinate $x-Vt$. Here, the two-hump solution travels with
velocity $V$ and is symmetric to the point $X(t)=Vt$. It should be noted
that due to the behaviour of the first and second moments, a connection of
the form $\langle x(t)\rangle_{V}\propto V\langle x^2(t)\rangle_0$ (see, e.g.,
Ref. \cite{report,eli}) between the first moment in the presence of the
constant drift $V$ to the second moment in absence of this drift, does not
exist, in contrast to the subdiffusive case \cite{mebakla}.

Similarly, for the Ornstein-Uhlenbeck potential $\Phi(x)=\frac{1}{2}m\omega^2
x^2$ which exerts the linear restoring force $F(x)=m\omega^2x$, the first
moment shows the exponential relaxation
\begin{equation}
\langle x(t)\rangle=x_0e^{-\omega^2t/\gamma}
\end{equation}
which contrasts the Mittag-Leffler patterns recovered in the
subdiffusive model in Refs. \cite{gcke} as well as the superdiffusive
fractional KKE presented in Ref. \cite{basil}.
The second moment has the Laplace transform
\begin{equation}
\langle x^2(u)\rangle=\frac{x_0^2+2Ku^{\alpha-2}}{u+2\omega^2/\gamma}
\end{equation}
whose inversion leads to
\begin{equation}
\label{mlp}
\langle x^2(t)\rangle=x_0^2e^{-2\omega^2t/\gamma}+\frac{2Kt^{2-\alpha}}{
\Gamma(3-\alpha)}\,_1F_1\left(1;3-\alpha;-\frac{2\omega^2}{\gamma}t
\right)
\end{equation}
which combines the exponential relaxation of the initial condition, which
was already found for the first moment, with the confluent hypergeometric
function $_1F_1$. Note that the second term in Eq. (\ref{mlp}) is equal
to the expression $2Kt^{2-\alpha}E_{1,3-\alpha}\left(-2\omega^2t/\gamma
\right)=2Kt^{2-\alpha}\sum_{n=0}^{\infty}\left(-2\omega^2t/\gamma\right)
/\Gamma(3-\alpha+n)$ in which we used the generalised Mittag-Leffler function
$E_{1,3-\alpha}$ \cite{erdelyi}. The variance is
consequently given in terms of
\begin{equation}
\Delta x(t)^2=2Kt^{2-\alpha}E_{1,3-\alpha}\left(-2\omega^2t/\gamma\right)
\end{equation}
which interpolates between the freely diffusive behaviour (\ref{msd}) for
short times, and the long time power-law pattern
\begin{equation}
\Delta x(t)^2\sim \frac{K\gamma}{\omega^2}\frac{t^{1-\alpha}}{\Gamma(2-
\alpha)}.
\end{equation}
Accordingly, the mean squared displacement $\Delta x(t)^2$ {\em increases}
in the course of time even for long times. It is straightforward to show
in general that the solution of the fractional Fokker-Planck-Smoluchowski
equation (\ref{ffp}) does not have a stationary solution. This is exactly
what is expected from a L{\'e}vy walk whose main characteristic is the
continuous approximation to a L{\'e}vy stable distribution \cite{klablushle}.

Let us now address the problem of negativity \cite{igoradbasil} which was
encountered for the fractional KKE proposed in Ref. \cite{basil}
in the overdamped limit. Firstly, it is possible to show that for all
constant or linear forces, moments of any order are positive. This is due
to the fact that our model equation (\ref{fkk}) leads to (generalised)
Mittag-Leffler-type relaxation patterns which are known to be strictly
monotonically decreasing and which are everywhere positive \cite{erdelyi}.
In contrast, the model from Ref. \cite{basil} leads to Mittag-Leffler
indices between 1 and 2 and therefore to temporal oscillations which in
turn give rise to negative portions in the ``pdf'' for shorter times
\cite{igoradbasil}. The positivity of the solution of our fractional KKE
(\ref{fkk}) can be demonstrated also for an arbitrary external force field.
We know that the velocity pdf $P(v,t)$ is positive for all times
\cite{report,basil}, and
that the position-space pdf $P(x,t)$ is positive in the force-free case
\cite{abm}. As the force enters through the local drift term,
also the pdf in the presence of an arbitrary external force $F(x)$ is
positive, and consequently the joint pdf $P(x,v,t)$ is positive.

Formally, the fractional KKE (\ref{fkk}) corresponds to the ``L{\'e}vy
rambling'' model developed in Ref. \cite{gcke}
but with the force entering symmetrically to friction and diffusion
(i.e., $\langle v\rangle=-\eta v\tau^*+F(x)\tau^*/m$ in Eq. (62) of Ref.
\cite{gcke}). Thus,
Eq. (\ref{fkk}) corresponds to an approximation to the CTRW-L{\'e}vy walk
model in which
the particle undergoes continuous motion between scattering events which
change the velocity coordinate. Friction and force remain events
which enter through an effective time scale $\tau^*$, i.e., they correspond to
point-like interactions in the long time limit $t\gg\max\{\tau,\tau^*\}$. This
gives rise to the
fact that the spatial distribution does not reach an equilibrium state, or,
in other words, that space and time do not decouple in the underlying
equations (\ref{fkk}) and (\ref{ffp}) so that the variables $x$ and $t$
cannot be separated, also a typical property of CTRW-L{\'e}vy walks.

We have introduced a new fractional approach to the phase space
description of superdiffusive sub-ballistic transport processes. The
obtained fractional KKE leads to the Mittag-Leffler
relaxation of the velocity distribution towards the classical 
Maxwell-Boltzmann equilibrium. In contrast, the long-time or high-friction
limit of the spatial distribution does not possess a stationary
solution. The process is characterised by a combination of the classical
time evolution found for the first moment, such as the exponential relaxation
of the initial condition in the presence of a linear force field, with
a time-dependence which is governed by the fractional order $\alpha$, i.e.,
the ``memory strength''. The fractional KKE fulfils a generalised Einstein 
relation in velocity space. In coordinate space, no analogous generalisation
of the Einstein relation exists. These properties set the present approach
apart from previous fractional models and we therefore regard it a
distinguished candidate for the description of enhanced, sub-ballistic
transport in external force fields, and as complementary to the established
subdiffusive fractional KKE-description.

We call the associated stochastic process a differential L{\'e}vy walk in
reference to the fact that Eq. (\ref{fkk}) correctly describes lower order
moments of a CTRW-L{\'e}vy walk in an external force field and in phase
space. A similar investigation of such situations within the original CTRW
scheme would be very tedious. We expect that this study will contribute to
a better understanding of superdiffusive transport and its dynamical
foundation.

We thank Yossi Klafter for helpful discussions. RM acknowledges financial
support from the Deutsche Forschungsgemeinschaft (DFG) within the Emmy
Noether programme. IMS acknowledges financial assistance from the
Fonds der Chemischen Industrie.

\end{document}